# The Wireless Technology Landscape in the Manufacturing Industry:
# A Reality Check

*Xavier Vilajosana, Cristina Cano, Borja Martinez, Pere Tuset, Joan Melià, Ferran Adelantado*
*Wireless Networks Research Lab. Universitat Oberta de Catalunya. Catalonia, Spain*
*{xvilajosana,ccanobs,bmartinezh,peretuset,melia,ferranadelantado}@uoc.edu*

## 1. Introduction

An upcoming industrial IoT revolution, supposedly led by the introduction of embedded sensing and computing, seamless communication and massive data analytics within industrial processes [1], seems unquestionable today. Multiple technologies are being developed, and huge marketing efforts are being made to position solutions in this industrial landscape. However, we have observed that industrial wireless technologies are hardly being adopted by the manufacturing industry. In this article, we try to understand the reasons behind this current lack of wireless technologies adoption by means of conducting visits to the manufacturing industry and interviews with the maintenance and engineering teams in these industries. The manufacturing industry is very diverse and specialized, so we have tried to cover some of the most representative cases: the automotive sector, the pharmaceutical sector (blistering), machine-tool industries (both consumer and aerospace sectors) and robotics. We have analyzed the technology of their machinery, their application requirements and restrictions, and identified a list of obstacles for wireless technology adoption. The most immediate obstacles we have found are the need to strictly follow standards and certifications processes, as well as their prudence. But the less obvious and perhaps even more limiting obstacles are their apparent lack of concern regarding low energy consumption or cost which, in contrast, are believed to be of utmost importance by wireless researchers and practitioners. In this reality-check article, we analyze the causes of this different perception, we identify these obstacles and devise complementary paths to make wireless adoption by the industrial manufacturing sector a reality in the coming years.

## 2. Analyzed Industries and Observations

We have visited industries in four different sectors to gather real information from actual deployments and use cases. The gathered information has also been compared to the use cases identified by the IETF Detnet WG [2].

### 2.1 Automotive

Within the automotive industry there is a large set of specific machinery to manufacture and assemble the different parts of a vehicle. One significant example of such machinery are press lines. The manufacturing of large parts, mainly for vehicle bodywork, requires the use of various presses. Presses are long-lasting machinery with operational lifetimes of more than 50 years. Newest presses today are large distributed systems, composed of thousands of embedded devices that control motors and hydraulic pressure pumps. Communication to the system controller is conducted through Ethernet buses with industrial control protocols, such as Profinet. Legacy machines (some of them from the 60s) use industrial field buses, with proprietary technologies like the legacy Simatic S5 PLC. We have not observed any wireless link in the presses, even in the newest ones (installed in 2017 in the plant). Accessories or newer sensors are rarely added, but in such rare cases they are provided by the press vendor using the specific machine interfaces, mostly based on Programmable Logic Controllers (PLC) and wired industrial networks.

### 2.2 Blistering in Pharma

In any pharmaceutical industry the packaging process is complex and involves large infrastructures and assembly lines. Each line is composed by different chained machines that assemble the pills, control that the blisters are correctly filled and box them. Machines integrate a large number of sensors and cameras (for computer vision control) interfaced by the machine central controller or a remote SCADA system. Hard real-time control happens inside the machine through field buses or deterministic Ethernet. We observed that the pharmaceutical plant is instrumented by an optical fiber-based double industrial Ethernet network to achieve redundancy. During our interviews, wireless technologies were only mentioned as an alternative to have redundant thermostats for the industrial HVAC (Heating, Ventilating and Air Conditioning). Similarly to the automotive industry, accessories are typically provided by the machinery vendors and use common PLC interfaces through wired industrial networks and protocols to achieve interoperability.

**2.3 Large Machinery-Tool Manufacturing**

Analogously to the automotive sector, the manufacturing of large machines involves complex production lines, typically constituted by custom-made machine-tools and robots that are and built in an ad-hoc manner to address one specific task in the production/assembly line. We visited an industry that develops machine-tools to manufacture airplane parts. These machines are large blocks (with a longitude in the order of 50 meters) with articulated arms and transportation platforms operated by hundreds of PLC systems. Communication between subsystems is performed through fiber-optics or deterministic Ethernet networks running industrial communication protocols, such as PROFINET, POWERLINK, EtherCAT or SERCOS. SCADA systems are used in the backend, with the newest ones offered as a cloud service instead of embedded in a given computer. We also visited another industry that manufactures metalforming machinery for the consumer sector. These machines are large distributed systems that bend metal to, for example, manufacture cocking utilities or vehicle rims. Cameras, pressure and temperature sensors are massively used to control the process. All the subsystems are interconnected through Ethernet cables and information is transported with MODBUS over TCP/IP. Real-time controllers use dedicated microcontroller buses such as PCI Express or CAN.

**2.4 Industrial Robotics**

Industrial robots are automated, programmable and usually articulated. The head or extreme of the arm can be adapted to different applications such as soldering or assembling. We visited a major articulated industrial robot manufacturer and analyzed the most advanced robots under development. They are composed of numerous distributed embedded controllers networked through field buses or industrial Ethernet. They use a large set of sensors and encoders to track movement and position. Motor control loops are handled by different dedicated control units. Hard real-time is handled by on-board buses such as CAN. Wireless communications are not used for the internal communication and control subsystem. In our visit, we found interest in using wireless communication to stream camera images from rotating parts in the extremes of the arm.

# 3. Identified Obstacles for Wireless Adoption

The most shocking fact of what we have observed is that wireless technologies are not widely adopted by the manufacturing industry nowadays. In this section we discuss the reasons we have identified behind this lack of adoption.

**1) Over-dimensioning**

We observed that, despite the application requirements imposed in certain subsystems are low, solutions based on fiber-optics providing nanosecond latency are used. This indiscriminate use of ultra-reliable wired technologies, despite their features are not being fully exploited, seems to be the norm. In general, ultra-reliable technologies are used even when not needed. The reasons we infer with respect to this fact are:

I) The *cost of communication* is insignificant compared to the cost of the industrial equipment. When considering large machinery, like a press line, or an aeroplane wing manufacturing machine, the cost of cabling and using the most ultra-reliable communication solution is insignificant compared to the whole machine. So, little concern is given to the cost of communication, which is in clear opposition to the design requirements addressed by researchers and practitioners that consider reduction of cost of utmost importance.

II) *Low power operation* is not of concern. We observed that all devices that connect to, or are part of the machinery or are already powered to a source of energy, removing the need for low power communication.

**2) Resistance to change**

We have observed a strict resistance to the adoption of new <u>substitute</u> communication technologies. Even when a communication technology fulfills the goal it is usually not changed. The reasons we can guess are:

I) The *cost of ownership* imposes resistance to adopt substitute technologies. When a candidate technology does not provide any perceived advantage to what is used, the effort of getting into it limits its adoption.

II) The *cost of implantation* and/or replacement. Introducing wireless technologies (as any other) require technical interventions that may impact on the production lines or production lifecycles, hence, manufacturers are very prudent and only drive interventions when is strictly needed.

III) *Limited support to legacy industrial standards*. Despite the large efforts conducted by the standardization bodies to interconnect "IoT" wireless devices to the Internet in the last years, industry perceives a lack of support of application-level and transport protocols addressing industrial technologies. The IETF CoAP, MQTT-S and other efforts have addressed the connection of constrained devices to Internet services, but we have seen almost no effort to support industrial transport and application protocols on top of wireless technologies. Industrial SCADA systems expose interfaces based on HTTP, OPC-UA, MODBUS, PROFINET but mostly not CoAP or MQTT yet.

IV) We have observed that when an industry uses a particular technology, and that technology performs as it is required, manufacturers stick to it. Even further, manufacturers stick to particular vendors because they value the service and the confidence that the solution will work in the long term.

### 3) Wireless performance is perceived to be poor

Despite several standards have emerged to provide reliable wireless communication (e.g PROFINET over 802.11 PCT, WirelessHART. IEEE802.15.4-TSCH, IETF 6TiSCH, DECT-ULE), we have seen them very rarely in the visited industries. Wireless is perceived as a non-reliable technology. We derived the following reasons:

I) Contention-access based technologies such as Zigbee have perhaps influenced the belief that wireless cannot be used for reliable communications.

II) We noted that the usual requirements for the type of control that is required [2] is quite far from the few milliseconds latency that can be guaranteed with deterministic wireless networks [4].

III) It is known that wireless networks performance is, in the best case, similar to an equivalent wired solution performance. The key advantages of wireless are in the operative side since cabling is not needed, however, this is not fully perceived as a key need for most of the use cases as described above.

### 4) Perception of an immature wireless market

There is a clear perception of quick obsolescence and market fragmentation of wireless technologies. Some of the inferred reasons are:

I) There is a belief that wireless technologies are very fragmented and that there is not a "de-facto" standard that can be adopted with confidence. This can be explained by the perceived youth of the current wireless communication market, where several new standards and proprietary solutions are emerging every year.

II) For the same reasons, there is also the belief that wireless technologies have quicker obsolescence than industrial wired standards. This is perceived as a risk by the industry, considering that machinery has a long operative lifetime, sometimes in the order of 30 years.

### 5) Environmental barriers

The environmental characteristics of an industrial scenario are quite challenging for wireless communications. Although pioneering studies demonstrate high levels of reliability for WirelessHART and pre-6TiSCH networks in real industrial settings [5], we have observed distrust in wireless technologies. This is specially relevant in large metallic machines or environments where concrete walls confine parts of the machinery.

## 4. Potential Paths for Wireless Adoption

In this section we devise some potential paths to follow in order to address the obstacles identified before.

### 1) Understand industrial requirements

Perhaps one of the most surprising identified obstacles for wireless adoption results from the apparent lack of concern for low energy consumption and cost of industrial communication devices, although these two aspects are seen by researchers and wireless practitioners as highly important to guarantee adoption by the industry. Therefore, one of the most important goals is to improve our understanding of their requirements. We believe that without increased communication channels with the industry, the adoption will continue to be just anecdotal.

### 2) Increase added value

We perceived a lack of appreciation of the value that can provide wireless technologies. We believe that this can be enhanced through the following paths.

*Fill the missing gaps today:* By identifying unresolved problems or by augmenting existing machinery functionalities, wireless can start building trust and foster adoption. Maintenance departments are key allies for wireless technologies as their deployment dramatically simplifies their task. Addressing support of industrial application and transport standards on top of wireless PHY and MAC technologies may foster adoption.

*Integration of wireless in machinery*: By working together with the machinery manufacturers we could then guarantee that standard industrial wireless solutions used by the industry do not quickly become obsolete. Integrating industrial wireless solutions in the machines from the vendors they already know strengthens trust, and allows for a reduction the cost of adoption and implantation.

*Reduce the cost of ownership:* The cost of ownership is a clear factor against change. Further efforts on standardization and transparent solutions to the application are needed. Efforts must be taken to simplify knowledge transfer and formation to the industry.  Also by adapting industrial well-established technologies over different wireless technologies so adoption is simplified.

### 3) Improve the perception about the technology
Since wireless technologies are perceived negatively, and over-dimensioning is the norm, technology perception improvement can be addressed by working on specific use-case deployments. In particular, those that can only be addressed by wireless technologies may be more appropriate. One such example is predictive maintenance of rotors. Features like temperature of the rotor's surface, directly linked to the motor's performance, is currently approximated with thermal cameras. However, embedded temperature sensors powered by battery-less RFID tags or small 6TiSCH networks may reliably predict performance issues, reducing maintenance downtimes and production interruptions. This can help the manufacturing industry to realise the benefits of wireless and create trust on the technology.

### 4) Strong standardization
Huge efforts are done towards Internet integration (IP-Enabled) of wireless technologies, which may now look on how to support the widely adopted industrial standards over IP (e.g Modbus, Profinet, etc..).  For instance, promoting the wireless standards already adopted in other industrial markets or sectors through standardization bodies alliances, recent standards, and consolidating wider specifications covering multiple verticals/sectors, may help to fill the missing gaps and to join industrial requirements. This will provide a vision of continuity of the technology, and hence, reducing the quick obsolescence perception.

### 5) Study the adequacy of more powerful wireless technologies
It seemed clear until now that low-data rate technologies were well-suited for other industrial scenarios such as metering, infrastructure monitoring, etc. However, to address low-latency control loops, to improve the current negative perception of performance, as well as to address industry preference for reliability via over-provisioning, higher capacity and more reliable wireless technologies compared with the traditional low-power IoT can benefit wireless adoption in a substantial manner.

## 5. Final Remarks
In this paper, we have presented the results of a reality check of wireless technologies adoption in industrial environments. We have analyzed the use level of these technologies in five different sectors and identified key challenges and needs. We have also identified strategies for the wireless technologies industry to channel these needs and shape future wireless technologies. Ultimately, this should help to materialize the wireless technology adoption in industrial environments, that is taking too long to happen.

## Acknowledgements

Work supported by the Spanish Ministry of Economy and the FEDER fund under SINERGIA (TEC2015-71303-R).